# Effective Mass in Bose-Einstein Condensation in the Bound State and Phonon Propagation in the Unbound States

Asis. K. Bandyopadhyay[1], Atrayee Bhattacharya[2], Kamal Choudhary[3], and Santanu Das[4]

1) GCECT, W. B. University of Technology, Kolkata-700010, India

2) Department of Applied Mathematics with Oceanology and Computer Programming, Vidyasagar University, Midnapore, 721102 India.

3) Department of Materials Science, University of Florida, Gainesville, FL 32611, USA.

4) Department of Ceramic Engineering, Indian Institute of Technology (BHU), Varanasi, 221005 India.

*Corresponding Author*: **asisbanerjee1000@yahoo.co.in**

**Abstract:**

The dark and bright solitons in different systems are already known in Klein-Gordon lattice. Instead of an external driving force, if the intrinsic field is only considered, then the modal dynamics for small oscillations could be characterized by the bound state in a limited range of frequency, revealed via associated Legendre polynomial. Bose Einstein condensation takes place around bosonic particles having different wave functions within the bound states in the temperature region $T = 0$ to $T_c$ having implication for the effective mass of the system. The pairing and interplay between the dark and bright solitons also occur with their effect on the condensation. This effective mass is calculated via statistical mechanics route by two-part partition function that also gives an indication for the transition temperature. The disappearance of the bound state after a critical frequency, or equivalently, after a critical temperature, gives rise to quasi-particles or phonons in the unbound states that propagate through the domains.

## I. **Introduction**:

The most outstanding experimental discovery in recent times is Bose-Einstein condensation (BEC) in 1995 **[1]** that has triggered both theoretical and experimental works on this fascinating topic of research. The reasons for these vigorous activities are: a) it gives an opportunity to open a new window for a macroscopic view of quantum mechanics, and b) it makes such studies most interesting in the field of matter-wave relations. On the latter issue, experiments show nonlinear excitation within the wave that is known as solitons. The properties of such quasi-particles, i.e. both bright and dark solitons (henceforth called bosonic particles, or simply particles, as in the BEC literature), in a condensate would allow us to manipulate them in periodic or other potential. There is some evidence for the formation of both dark and bright solitons in the condensate (see later), but the concept of bound state still remains somewhat illusive, despite a lot of activities on nonlinear optical systems that are important for many devices. This gives us motivation to study bound state with a connection to BEC. Further, we find out both lower and upper bounds in relation to frequency and the extent of condensation in the bound state. Although we use some data on lithium niobate ferroelectrics for the general theoretical study, it can also be extended to other relevant systems, e.g. in magnon system with two-well Landau potential **[2]**. This effort could be made through Klein-Gordon (K-G) equation which on perturbation gives rise to nonlinear Schrodinger equation (NLSE) that is a variant of Gross-Pitaevskii equation (GPE), which is popular and commonly used for studying BEC.

K-G equation was developed earlier for the ferroelectric system **[3]** (see the references therein for the other important investigations) through variational principle taking various relevant energies in the Landau-Ginzburg (L-G) potential in the continuum Hamiltonian. Recently, this was extended to nonlinear optical materials, such as ferroelectrics and metamaterials, using one dimensional array of domains to probe localization for intrinsic localized modes **[4]**. Very recently, a perturbation on the continuum K-G model giving rise to NLSE **[5]** showed both bright and dark solitons with 'discrete energy levels', estimated via hypergeometric function. This dark soliton with low energy and lower velocity is not visible as it is part of the complex solution indicating the presence of an energy-gap. In this communication, for these solitons, our main focus lies on the bound state in the low frequency limit in the context of BEC. A critical frequency is also shown beyond which the phonon takes over in the unbound states in the high frequency regime.

There are excellent reviews on BEC in the vast ocean of literature, notably Ref **[1]**. Regarding solitons, it is simply impossible to write about them in detail, but among many important books on the subject, it is worth mentioning about the book by Dauxois and Peyard and also about a very popular book on optical solitons by Kivshar and Agarwal **[6]**. Recent theoretical investigations on discrete solitons need to be mentioned, as in Ref. **[7]** that contains many important references on solitons in general. Now, in a non-exhaustive literature search, in the context of BEC some important references are given here. The dynamical generation and control of bright solitons in BEC of $^7$Li were done by Strecker et al **[8,9]** by utilizing a Feshbach resonance to switch from repulsive to attractive interactions. The existence of gap solitons in BEC was shown by Eiermann et al **[10]**. The experimental evidence of dark solitons in matter-wave BEC was first studied by Burger et al **[11]**, and through phase imprint method or phase engineering by Denschlag et al **[12]**, and Anderson et al **[13]** also studied dark solitons in BEC. In this context, a recent review on dark solitons in atomic BEC by Frantzeskakis **[14]** also needs to be mentioned; this review contains many important references. In a slightly unrelated work, both dark and bright solitons were studied in a hard-core boson system by a non-GPE type of equation using spin-coherent state averages by Balakrishnan et al **[15]**, wherein both the solitons played their role below and above acoustic velocity. Ohberg and Santos **[16]** made an interesting study in a two components system on soliton-soliton bound pair wherein dark soliton can be transferred from one component to the other at the domain wall when it exceeds a critical velocity. All these investigations motivate us for taking a further look at the bound state and the effective mass in the context of BEC as well as frequency-temperature relation.

Our intention is to show that at low temperature there is BEC formation within a small frequency regime that is described in terms of frequency as well as temperature in the bound state and also phonon propagation at or near room temperature in the unbound state. The latter gives a theoretical confirmation on 'acoustic memory'. To meet our objective, first of all, let us look at the potential and the bound states.

The paper is organized as follows: In **Section II**, after describing the potential and both the spatial and spatio-temporal equations, we deal with the linearization problem and then we arrive at associated Legendre polynomial that gives us indication for the boundary limits of the bound state. In **Section II**, we use partition functions involving Landau energy and follow a standard statistical mechanics route to arrive at effective mass and also the frequency-temperature relation is worked out to be able to relate it to the BEC formation. In **Section III**, we describe different wave functions to indicate dark-bright soliton interaction and on the BEC formation at different frequencies within the bound state. Also, we show that after a critical value of frequency, the phonons start propagating through the domains within the system in the unbound state. Conclusions are given in the **Section IV**.

**II. Theoretical Development:**

The free energy density for the order parameter ($P$) can be written in Landau-Ginzburg form: $G = -(\alpha_1/2)P^2 + (\alpha_2/4)P^4$. Where, α's are Landau coefficients and here $P$ is also a function of space. Now, $\frac{\partial G}{\partial P} = E = -\alpha_1 P + \alpha_2 P^3$. Here, $E$ is the intrinsic field (dimensional). The relevant values in the context of a ferroelectric system is given in Ref. **[4,5]**. The 2$^{nd}$ derivative of Landau energy is:

$$g''(P) = E/E_c = -\bar{\alpha}_1 + 3\bar{\alpha}_2 P^2 \tag{1}$$

Here, the non-dimensional field is external, where, $E_c$ is the switching field in kV/cm, and the Landau coefficients are non-dimensional. Let us consider an idealized one-dimensional array of $N$ identical rectangular domains along the $x$ direction. Between the neighboring domains, there is domain wall and nearest neighbor coupling ($K$) is considered. For the mode dynamics of the extended modes, nonlinear K-G equation relating $P$ against space ($x$) and time ($t$) with a non-dimensional driving field ($E_0$) is **[3,5]**:

$$\frac{\partial^2 P}{\partial t^2} - \bar{K}\frac{\partial^2 P}{\partial x^2} - \bar{\alpha}_1 P + \bar{\alpha}_2 P^3 - E_0 = 0 \tag{2}$$

Here, $E_0$ is the external driving force. K-G equation is well-known in mathematical physics that exhibits a variety of interesting properties with applications in different physical systems **[3-5]**. Although K-G equation being classical is useful for both dark and bright discrete breathers **[4]**, as a passing reference, it may be mentioned that the concerned Hamiltonian upon quantization throws light on the quantum localization **[17]**. Next, let us go for the solutions: In the continuum limit, let $P$ be the solution of Eq. (2) that is replaced by $P = P(x) + f(x,t)$. Here, $P(x)$ and $f(x,t)$ are the functions of $x$ and $(x,t)$ respectively. From physics point of view this combination describes a periodic kink which, a priori, can experience the presence of phonons about its centre of mass regardless of its dynamical property **[18]**. Thus, the resulting eigenvalue equation will be governed by a linearized problem. Let us write the space dependent equation:

$$-\bar{K}\frac{\partial^2 P(x)}{\partial x^2} - \bar{\alpha}_1 P(x) + \bar{\alpha}_2 P^3(x) = 0 \qquad (3)$$

As said earlier, our main focus area is the bound state with a connection to BEC. Here, we use associated Legendre polynomial (ALP) to reveal a much richer physics by showing stable "upper and lower bounds". To note that in a different context for light-induced waveguide, Segev et al. **[19]** used ALP function for modal composition of incoherent spatial solitons in nonlinear Kerr medium. Next, let us go for the spatio-temporal equation:

$$\frac{\partial^2 f(x,t)}{\partial t^2} - \bar{K}\frac{\partial^2 f(x,t)}{\partial x^2} + g''(P)f(x,t) - E_0 = 0 \qquad (4)$$

Here, $P(x)$ given in Eq. (3) is the static single 'kink' solution (not discussed here) with a form: $P(x) = \tanh qx$, where, $q = \sqrt{\bar{\alpha}_1/(2\bar{K})}$. Using this form of $P(x)$ and taking $\bar{\alpha}_1 \approx \bar{\alpha}_2$ in Eq. (1), we have

$$g''(P) = -\bar{\alpha}_1 + 3\bar{\alpha}_2 \tanh^2 qx = 2\bar{\alpha}_1\left(1-(3/2)\sec h^2 qx\right) \qquad (5)$$

$g''(P)$ varies mainly in the region of the kink centre (assumed to be at $x=0$) and approaches a constant value (taken to be unity) far from the kink centre, and also $g''(P) < 0$ at $x = 0$. For small oscillations ($E_0 \approx 0$), i.e. the present approach is based on 'intrinsic field' and it hardly depends on the external force. Now, $f(x,t)$ is written as: $f(x,t) = \psi(x)e^{-i\omega t}$ ($\omega$ = angular frequency). Substituting $f(x,t)$ and Equ. (5) in Equ. (4), we have Equ. (6) with $E_0 = 0$. Equ. (6) is expressed in two ways, such as ALP and operator, as shown later in equation (27). We think that both bright and dark solitons exist in the bound state. Operator is used to operate on bright and dark solitons to get respective frequency. Similar linearization equations (Eq. 23, Eq. 24) have been lucidly deduced by Sugiyama [20] in the context of kink-antikink collisions, but here our aim is quit different and also we use Landau potential whose coefficients give us indication on frequency or energy, as explained above in Section 1. Now, the eigenvalue equation can be written as:

$$\left(\bar{K}\frac{\partial^2}{\partial x^2} + 3\bar{\alpha}_1 \sec h^2 qx\right)\psi = X\psi \qquad (6)$$

$X$ is the eigenvalue of the system defined as: $X = (2\bar{\alpha}_1 - \omega^2)$. Eq. (6) is identical with the Schrodinger equation for a particle moving in one-dimensional potential well ($g''(P)$). This is considered as a variant of GPE if we add a vector potential. So, the bound and unbound states can be observed for this potential. For bosonic particles, let us introduce ALP to estimate the frequency of non-degenerate states with different wave functions. For soliton dynamics, in the low frequency range, the bound state will predominate, when condensation takes place and in the higher range, the phonon mode takes over in the unbound state, and here these are the main issues. Let us introduced a new variable: $z = \tanh qx$, then Eq. (6) becomes

$$(1-z^2)\frac{\partial^2 \psi}{\partial z^2} - 2z\frac{\partial \psi}{\partial z} + \left(n(n+1) - \frac{m^2}{1-z^2}\right)\psi = 0 \tag{7}$$

Where, $m^2 = 2X/\bar{\alpha}_1 = 4 - (2\omega^2/\bar{\alpha}_1)$. The solution of Eq. (7) is: $\psi = p_2^m(z)$. In the bound state, $\omega$ is denoted as $\omega_b$ and wave function $\psi$ as $\psi_b$. If $0 \leq \omega_b \leq \sqrt{2\bar{\alpha}_1}$, then 'm' is real and ALP is only valid if $n=2$. The existence of different states is considered in the bound state in this limited range of frequency or energy. Here, all our solutions are 'real' and 'stable' in terms of interplay between the mode index and the frequency. When the bound state frequency, $\omega_b > \sqrt{2\bar{\alpha}_1}$, 'm' will be imaginary. Under this condition, ALP is not valid, i.e. the bound state disappears. This is considered as a "critical" limit of frequency for the 'upper bound'. The impact of this critical limit on condensation is shown at the end of Section 3.1.

At this stage, before discussing our focus issue (BEC) within a smaller range of frequency in the bound state, let us also talk about an important aspect of BEC, i.e. the effective mass ($m^*$).

The Eq. (4) can also be derived from the Hamiltonian of the system that is written as:

$$H = \int_0^L (dx/a)\left[\frac{f_{i+1}^2}{2} + \frac{\bar{K}}{2}(f_{i+1} - f_i)^2 + g''(P)\frac{f_{i+1}^2}{2}\right] \tag{8}$$

$$H = \int_0^L dx H_1 = NH_1 \tag{9}$$

Where, the Hamiltonian can be written as:

$$H_1 = \frac{1}{a}\left[\frac{f_{i+1}^2}{2} + \frac{\bar{K}}{2}(f_{i+1} - f_i)^2 + g''(P)\frac{f_{i+1}^2}{2}\right] \qquad (10)$$

Where the lattice spacing 'a' has a value of 0.5 nm for lithium niobate. Dividing sample length 'L' into N segments, an explicit expression for partition function follows from the Hamiltonian $H_1$ as given below:

$$Z = \frac{V^N}{h^N N!}\int \prod_i^N \delta f_i \delta f_{i+1} \exp(\beta H_1) = Z_1 Z_2 \qquad (11)$$

Where, $\beta = 1/(k_B T)$ and the expression for the first part of partition function ($Z_1$) is shown as:

$$Z_1 = \frac{V^N}{h^N N!}\left(\frac{2\pi a k_B T}{\bar{K}}\right)^{\frac{N}{2}} \pi_i^N \int_{-\infty}^{\infty} \delta \dot{f}_{i+1} \exp[-\beta \dot{f}_{i+1}^2/(2a)] = \frac{V^N}{\lambda_T^N N!} \qquad (12)$$

Where $k_B$ is Boltzmann constant and $\lambda_T$ is the thermal wavelength defined as:

$$\lambda_T = \left(\frac{\hbar\sqrt{\bar{K}}}{2\pi a k_B T}\right) \qquad (13)$$

And the second part of the partition function can be expressed as:

$$Z_2 = \left(\frac{2\pi a k_B T}{\bar{K}}\right)^{\frac{N}{2}} \pi_i^N \int_{-\infty}^{\infty} \delta f_{i+1} \exp[-(\beta/a) F(f_{i+1}, f_i)] \qquad (14)$$

With the following functional form that is expressed as:

$$F(f_{i+1}, f_i) = \frac{\bar{K}}{2}(f_{i+1} - f_i)^2 + g''(P)\frac{f_{i+1}^2}{2} \qquad (15)$$

The $\phi_n$ being the distribution function for field amplitude $f$ and delta function is defined as:

$$\delta(f_1 - f_1') = \sum_n \phi_n^*(f_1')\phi_n(f_1) \qquad (16)$$

With $f_1' = f(0), f_1 = f(L)$. So, we can define Equ. (14) as:

$$Z_2 = \sum_n \left(\bar{K}/(2\pi a k_B T)\right)^{N/2} \int_{-\infty}^{\infty} df_1...df_N \phi_n^*(f_1') \exp\left((-\beta/a)F(f_1,f_n)\right)...\phi_n(f_1)\exp\left((-\beta/a)F(f_2,f_1)\right) \quad (17)$$

This expression for $Z_2$ can be directly evaluated if $\phi_n$ are eigenfunction of the transfer operator that is expressed as:

$$\left(\bar{K}/(2\pi a k_B T)\right)^{1/2} \int_{-\infty}^{\infty} \delta f_i \exp\left((-\beta/a)F(f_{i+1},f_i)\right)\phi_n(f_i) = \exp\left((-\beta/a)E_n\right)\phi_n(f_{i+1}) \quad (18)$$

so that the second part of the partition function can finally be expressed as:

$$Z_2 = \sum_n \left(-\beta(L/a)E_n\right) \quad (19)$$

For a thermodynamic system in which $(L/a) \to \infty$, only ground state contribution is considered, and the free energy per unit length $= \left(-(k_B T)/L\right)\ln Z_2$. Applying the method of Ref. **[21,22]** we have

$$\left(\bar{K}/(2\pi a k_B T)\right)^{1/2} \int_{-\infty}^{\infty} \delta f_i \exp\left((-\beta/a)F(f_{i+1},f_i)\right)\phi_n(f_i) = \exp\left((-\beta/a)H_{eff}\right)\phi_n(f_{i+1}) \quad (20)$$

With the effective Hamiltonian expressed as:

$$H_{eff} = -a^2/(2\beta^2 \bar{K})\frac{\partial^2}{\partial f_{i+1}^2} + g''(P)(f_{i+1}^2/2) + C_0 \quad (21)$$

Where $C_0 = i\pi$ = constant, that appeared due to phase change in conversion of $\phi_n(f_i)$ to $\phi_n(f_{i+1})$. Substituting Eq. (20) in Eq. (18) we have

$$\left(-a^2/(2\beta^2 \bar{K})\frac{\partial^2}{\partial f_{i+1}^2} + g''(P)(f_{i+1}^2/2) + C_0\right)\phi_n(f) = E_n\phi_n(f) \quad (22)$$

### III. Results and Discussion:

### 3.1 Bose-Einstein Condensation:

From Eq. (7), the 'lower bound' of the non-degenerate state is at $\omega_b = 0$ for which the wave function ($\psi_b$) with translation symmetry gives rise to Goldstone mode (GM) for $m = 2$:

$$\psi_b = P_2^2(z) = 3\sec h^2 qx \tag{23}$$

To note that *only* bright solitons predominate at zero frequency. This wave function for bosonic particles is not shown here, as it simply shows a typical Gaussian band. The wave functions for other symmetries of GM were not worked out to remain within our main focus on the bound state and BEC formation. As the frequency increases to: $\omega_b > \sqrt{(3\bar{\alpha}_1)/2}$, the system starts showing polarization within a band of $m = \pm 1$, whose wave functions are:

$$\psi_b = P_2^1(z) = 3\tanh qx.\sec hqx \tag{24}$$

$$\psi_b = P_2^{-1}(z) = -(1/6)\tanh qx.\sec hqx \tag{25}$$

To note that both dark and bright solitons exist, as clearly evident from the above two equations, and a pairing or coupling seems to have started in the system. A small number of particles become polarized in opposite directions with the above value of eigenfrequency. These wave functions when multiplied by the exponential (read, temporal) term give rise to $f(x,t)$ or simply $f_b$. This is shown in **Fig. 1a** and **Fig. 1b** respectively corresponding to the above two spatial wave functions indicating that this behavior manifests in both + and – directions starting at zero. Finally, the wave function for the 'upper bound' of the non-degenerate state with the limiting frequency $\omega_b = \sqrt{2\bar{\alpha}_1}$ is:

$$\psi_b = (1/2)(3\tanh^2 qx - 1) \tag{26}$$

To note that there is no bright solitons or *sech* term and dark solitons have completely taken over, before going to the unbound state.

After clearly observing both lower and upper bounds [as in Eq. (23) and (26)], we are inclined to extend our results for both dark and bright solitons through a compact operator that is designed to evaluate the frequency for a chosen wave function. Following Eq. (6), the eigenvalue equation can be written as: $L\psi_b = X\psi_b$. Here, $X = (2\bar{\alpha}_1 - \omega^2)$ is the eigenvalue of the wave function $\psi_b$ operated by the operator $L$ as:

$$L = \frac{X}{2}(L_+ L_- + L_- L_+) \qquad (27)$$

Where, $L_- = \sqrt{\bar{K}/X} \left[ \sqrt{(3\bar{\alpha}_1/\bar{K})} \sec hqx + i\frac{\partial}{\partial x} \right]$ and

$L_+ = \sqrt{\bar{K}/X} \left[ \sqrt{(3\bar{\alpha}_1/\bar{K})} \sec hqx - i\frac{\partial}{\partial x} \right]$.

With this compact form, let us deal with the solitons as a preamble to our discussion on the BEC formation and number density. For static dark soliton, the above eigenvalue equation contributes to the wave function with translation symmetry as: $\psi_b = \sqrt{N_0} \tanh qx$. This spatial relation is valid in the bound state within the corresponding frequency. Here, $N_0$ is the total number of bosonic particles. From the above operator (Eq. 27), this idea leads to the dispersion relation for small oscillations:

$$\omega_b^2 = 2\bar{\alpha}_1 \tanh^2 qx \qquad (28)$$

This frequency relation is useful for explaining the formation of BEC. The solution of Eq. (4) leads to the spatio-temporal wave function as: $f_b(x,t) = \sqrt{N_0} \tanh qx.e^{-i\omega t}$. The dark soliton wave function and the frequency (Eq. 28) are plotted against spatial dimension in **Fig. 2**. For BEC formation around bosonic particles in a very low frequency range within a smaller spatial extent, it is noted from **Fig. 2** that the frequency curve touches the dark soliton wave function at this frequency range, and then it goes up again with the wave function curve. This could be considered as giving the signature of BEC that is of course subject to the experimental confirmation. It is also noted that the dip of frequency curve does not cover up the dark soliton wave function (but merely touches it). The underlying physics should tell us that here bright solitons dominate in the system, as also shown above from various wave functions. Let us now describe the bright soliton.

The bright soliton wave function is: $\Psi(x) = \sqrt{N_0} \operatorname{sech} qx$ using the above operator (Eq. 27) with the spatial dispersion relation:

$$\omega_b^2 = 2\bar{\alpha}_1 - 2\bar{\alpha}_1 \operatorname{sech}^2 qx - 1/2 = 2\bar{\alpha}_1 \tanh^2 qx - 1/2 \tag{29}$$

It is seen that the energy is lower than that of dark soliton. Now, the solution of equation (4) is: $f_b(x,t) = \sqrt{N_0} \operatorname{sech} qx \cdot e^{-i\omega t}$. Similar to **Fig. 2**, here we also plot bright soliton wave function and the frequency (Eq. 29) in **Fig. 3**. This shows that in a very small frequency range around zero, the bright soliton wave function indicates a maximum thereby proving the existence of only bright solitons at very low frequency. It was theoretically shown in Ref. **[5]** that bright solitons exist even at zero energy for the lithium niobate system. As shown below, bright soliton is totally condensed near or at zero frequency, since its energy is lower than that of dark soliton by (1/2), comparing Eq. (28) and (29). These figures seem to indicate a 'pairing' of two solitons, as the frequency starts increasing. It appears that there are various possibilities of conversion from dark to bright solitons (and vice-versa) and pairing within them could occur in the bound state. Having discussed both dark and bright solitons, and their 'interplay', let us now show the evidence of BEC.

It is considered that within the limited frequency range, the condensation takes place as per the dispersion relations. Due to bosonic particles within the bound state, their amplitude can be expressed by number density. In the context of matter-wave relation, as the frequency starts increasing, the dark solitons seem to ride on the back of bright solitons, which are totally condensed at zero frequency. Hence, some particles are considered as remaining outside the 'condensate'. The probability of finding these "free particles" is: $f_b^* f_b = N_0 - N$, and by taking the dark soliton wave functions we derive $N_0 \tanh^2 qx = f_b^* f_b = N_0 - N$. Where, $N$ is the condensed particle density. This is re-written as: $[1-(N/N_0)] = \tanh^2 qx$. Below BEC transition (for a finite temperature) using the dispersion relation (Eq. 28), the form of BEC relating number density and the frequency is:

$$(N/N_0) = 1 - (\omega_b^2 / 2\bar{\alpha}_1) \qquad (30)$$

This result might inspire the 'experimentalists' to work with larger molecules (multispecies) containing lithium **[23,24]** (see also Ref. **[25]** to **[27]**). As shown below, when the frequency increases the particles in the condensate decreases (Eq. 30). Here, the Landau coefficient (read, nonlinearity) assumes more significance. The number density of BEC is shown against frequency in **Fig. 4**. It is seen that at $\omega_b = \sqrt{2\bar{\alpha}_1}$, at a value of Landau coefficient of 353.42 (equivalently at 0.133 mole% niobium antisite defect at a switching field of 40kV/cm) **[5]**, $N/N_0$ goes to zero at upper bound. For 'quantum breathers' by quantizing discrete K-G Hamiltonian with Bosonic operators, a quantum pinning transition was observed at 40 kV/cm under periodic boundary condition in the 'phonon hopping coefficient' vs. impurity plot in lithium niobate **[17]**.

At Goldstone mode ($\omega_b = 0$), all the bright solitons get condensed (**Fig. 3**) in the bound state as per Eq. (23). For $\omega_b = \sqrt{\bar{\alpha}_1}$, $N = N_0/2$, which means that half of the number density of particles is condensed, i.e. the number of uncondensed particles (dark solitons) starts increasing. As the frequency in the system further increases to: $\omega_b = \sqrt{(3/2)\bar{\alpha}_1}$, $1/4^{th}$ of particles is condensed and $3/4^{th}$ of the number density are still in the bound state, also revealing polarized particles in the opposite directions at $m=\pm 1$, as per ALP formalism. This could be construed as the origin of polarization vis-à-vis bound state just below the upper bound. Finally, at criticality, $\omega_b = \sqrt{2\bar{\alpha}_1}$, the bound state totally disappears, and no particle can be condensed anymore in the system. Thus, in the context of BEC, the ALP formalism (Eq. 23 to 26) is quite noteworthy. It is worth mentioning that Ohberg and Santos **[16]** worked out a frequency range wherein soliton-soliton bound state seems to exist.

### 3.2 <u>Effective Mass</u>:

At this stage, after discussing our focus issue (BEC) within a smaller range of frequency in the bound state, let us also talk about another important aspect of BEC, i.e. the effective mass ($m^*$) **[28]**. The basic equations and all the relevant terms are explained at the end of **Section II** involving the standard treatment of statistical mechanics for partition function including our present formulation of free energy density and also frequency relations. The relation between quantum mechanical and statistical problem offers the role that time is exchanged for the distance times $i$ and $\hbar$ is replaced by $ak_BT$, as described by Scalapino et al **[22]**. The strong temperature dependence of an effective mass $m^*$ is defined by Krumhansl and Schrieffer **[29]** as:

$$\frac{a^2}{2\beta^2 \bar{K}}\left(\frac{\partial^2}{\partial f_{i+1}^2}\right) \equiv \frac{\hbar^2}{2m^*}\left(\frac{\partial^2}{\partial f_{i+1}^2}\right) \qquad (31)$$

so that we can write the expression for effective mass in relation to temperature as:

$$m^* = \frac{\hbar^2 \bar{K}}{\left(ak_BT\right)^2} \qquad (32)$$

At low temperature such as at T=0 K, the effective mass becomes infinity. At Goldstone mode, a very large value of effective mass ($m^*$) demands that the bosonic particles are condensed at the bottom of the bound state. At the upper edge of the bound state, the effective mass becomes 0.25 (compared to our assumption of mass to be equal to 1 in Ref. **[5]**) and the corresponding critical temperature ($T_c$) is thus $6.1 \times 10^{-4}$ K or, 0.061 micro-Kelvin. When T > $T_c$ the effective mass ($m^*$) becomes so small that the bosonic particles come out of the bound state and propagate as phonons through the unbound state. Here, the critical temperature is defined as:

$$T_c = \frac{2\hbar\sqrt{\bar{K}}}{ak_B} \qquad (33)$$

### 3.3 Temperature-Frequency Relation:

The relation between the frequency and temperature needs to be worked out. Ignoring $C_0$ from Eq. (21), we have made evaluation of the eigenvalue ($E_n$) for both bound and unbound states as:

$$E_n = \left(n+\frac{1}{2}\right)\frac{ak_BT}{\bar{K}}\left(g''(P)\right)^{1/2} = \left(n+\frac{1}{2}\right)\frac{ak_BT}{\bar{K}}(-\bar{\alpha}_1+3\bar{\alpha}_1 \tanh^2 qx)^{1/2} \qquad (34)$$

As we go from bound to unbound state, the temperature dependence of frequency from upper boundary of bound state to unbound state is defined as:

$$\omega_n \approx \left(n + \frac{1}{2}\right) \frac{a k_B T}{\hbar \bar{K}} (2\bar{\alpha}_1 \tanh^2 qx)^{1/2} \tag{35}$$

The spatial extension ($x$) between the lower and upper edges of the bound state is about $1.19 \times 10^{-7}$ m such that at the upper boundary of the bound state, the value of $\tanh^2 qx$ will be unity. At the upper boundary of the bound state, the temperature will be $T_c$ (Eq. 33) and frequency ($\omega_0$) will be $\sqrt{2\bar{\alpha}_1}$ for $\bar{K} = 4$. When $x > 1.19 \times 10^{-7}$ m, taking always $\tanh^2 qx = 1$, the Eq. (35) becomes

$$\omega_n = \left(n + \frac{1}{2}\right) \frac{a k_B T}{\hbar \bar{K}} \left(2\bar{\alpha}_1\right)^{1/2} \tag{36}$$

We assume that bound state, addressed as ground state ($n = 0$), leads to Eq. (36) as:

$$\omega_b = \frac{T}{T_c} \left(2\bar{\alpha}_1\right)^{1/2} \tag{37}$$

Eq. (37) and Eq. (30) give rise to the equation of BEC in the bound state relating number density and temperature as:

$$\frac{N}{N_0} = 1 - \frac{T^2}{T_c^2} \tag{38}$$

The frequency depends on the dimension of the sample. If we consider sample length $L$ then the Eq. (36) becomes:

$$\omega_n = \left(n + \frac{1}{2}\right) \frac{a k_B T}{L \hbar \bar{K}} \left(2\bar{\alpha}_1\right)^{1/2} \tag{39}$$

The above equation is useful in deriving a relation for memory and related issues that are offered by this novel material.

**IV. Conclusions:**

At low temperature, in a smaller frequency range, with very small oscillations both the lower and upper bounds of the bound state are derived through associated Legendre polynomial formalism. Bose-Einstein condensation is shown in this system in the realm of bound state that disappears after a critical frequency and thereafter no more condensation is possible in the system. Within the bound state, the 'interplay' between the dark and bright solitons gives rise to pairing or coupling between them that increases the dark solitons reducing the condensate density. Also, at or near room temperature with increasing frequency, the phonons start propagating through the domains within the system in the unbound state when scattering could be important.

**Acknowledgements:** The authors would like to thank Dimitri Frantzeskakis of Dept. of Physics at University of Athens (Greece) for many interesting suggestions. The authors would also like to thank Arthur McGurn of Dept. of Physics at Western Michigan University (USA), Larry Schulman of Dept. of Physics at Clarkson University (USA) for useful discussions and Andrew Sykes of Los Alamos National Laboratory (USA) for providing useful information.

**Figure Captions:**

**Fig. 1a:** The spacio-temporal wave function ($f_b(x,t)$) for the quasi-particle when $m = +1$. The behavior is seen to move towards positive direction starting at zero.

**Fig. 1b:** The spacio-temporal wave function ($f_b(x,t)$) for the quasi-particle when $m = -1$. The behavior is seen to move towards opposite direction starting at zero, compared to that in Fig. 1a.

**Fig. 2:** For small oscillations in the bound state, the frequency dispersion relation and the dark soliton wave function behavior for a coupling value= 4. The frequency curve is seen to touch the wave function for the dark soliton at zero frequency giving the signature of BEC. Below and above this coupling value, the frequency curve does not coincide with the dark soliton behavior.

**Fig. 3:** For small oscillations in the bound state, the frequency dispersion relation and the bright soliton wave function behavior for a coupling value= 4. The frequency curve is seen to fully cover up the wave function peak for the bright soliton indicating full condensation at Goldstone Mode frequency.

**Fig. 4:** The number density of BEC is shown against frequency. It is seen that at $\omega_b = \sqrt{2\bar{\alpha}_1}$ =26.59. This frequency is for Landau coefficient of 353.42, when the fraction of BEC condensation ($N/N_0$) goes to zero, and this frequency is considered as the 'critical limit' beyond which there is no more condensation of bosonic particles in the system, i.e. at or beyond the upper bound of the bound states, when phonon propagation starts in the unbound state.

*Appendix-A*: Having shown BEC in the lower regime of frequency in the bound state, it is natural to look for what happens in the unbound state. After the above critical limit of frequency, the switchover of bound → unbound state takes place and all the quasi-particles as 'phonons' will propagate in a given domain region in the system and through the domain wall. The phonons will propagate through the system with strong dispersion relation as:

$$\omega_P^2 = 2\bar{\alpha}_1 + k^2 \bar{K} q^2 \tag{A.1}$$

Where $k$ is the propagation constant and now $\omega_b$ has to be replaced by phonon frequency ($\omega_P$). It is convenient to use Eq. (A.1) in Eq. (6) that produces phonon solutions for $0 \leq k \leq \infty$ without using the normalized constant of $1/\sqrt{(3\pi)}$ as:

$$\psi_P = (1/\sqrt{1+k^2})\left(3\tan h^2 qx - 3ik \tanh qx - (1+k^2)\right)e^{ikqx} \tag{A.2}$$

This solution indicates propagation through one-dimensional array of domains in $x$ direction. The complex conjugate of the above function for waves that propagate through the system along the -$x$ direction is expressed as:

$$\psi_P^* = (1/\sqrt{1+k^2})\left(3\tan h^2 qx' + 3ik \tanh qx' - (1+k^2)\right)e^{-ikqx'} \tag{A.3}$$

In the wave function of phonons in the unbound state, if we put $k= 0$, we get the bound state solution at the "critical limit" (**Eq. 26**), which implies "localization" at lower energy that again indicates a richer physics. When phonons pass through the nonlinear optical medium, they encounter scattering in the vicinity of domain wall. Here, the Green function has a role to play in explaining this scattering in which phonons get shifted from one point to another point that would be the subject matter of future work in revealing the switching behavior in such systems.

**Figure 1a**

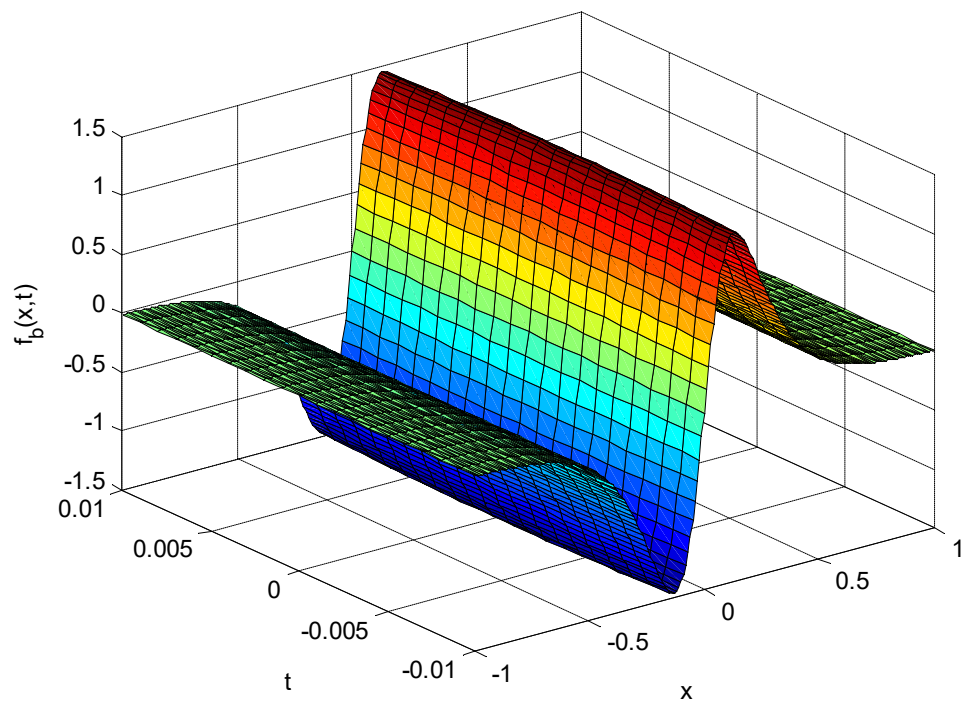

**Figure 1b**

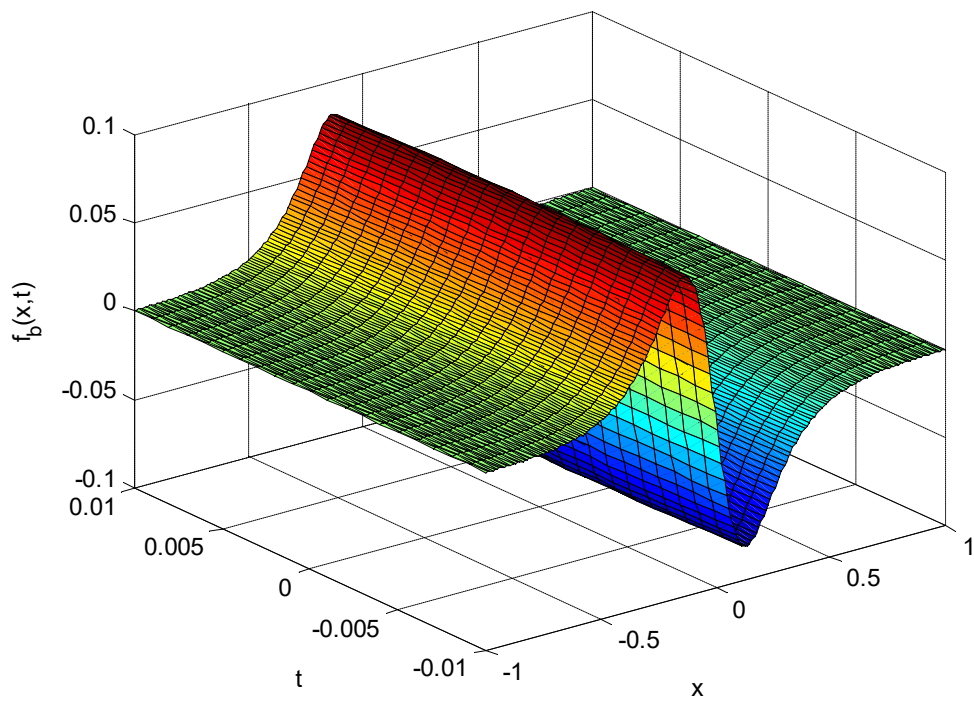

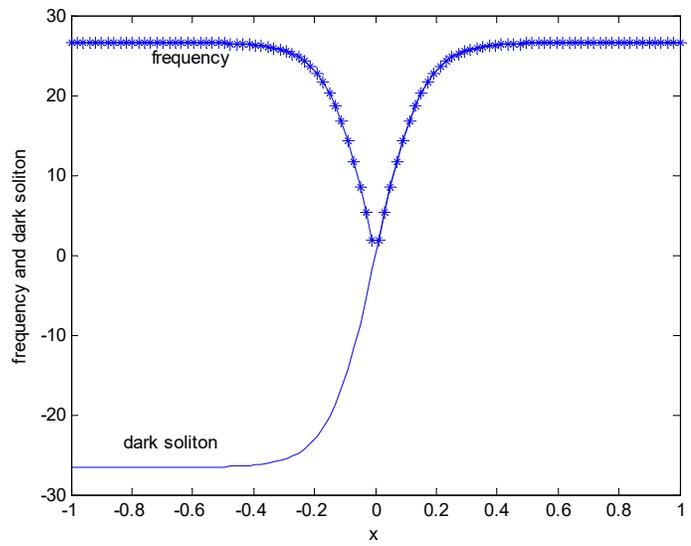

**Figure-2**

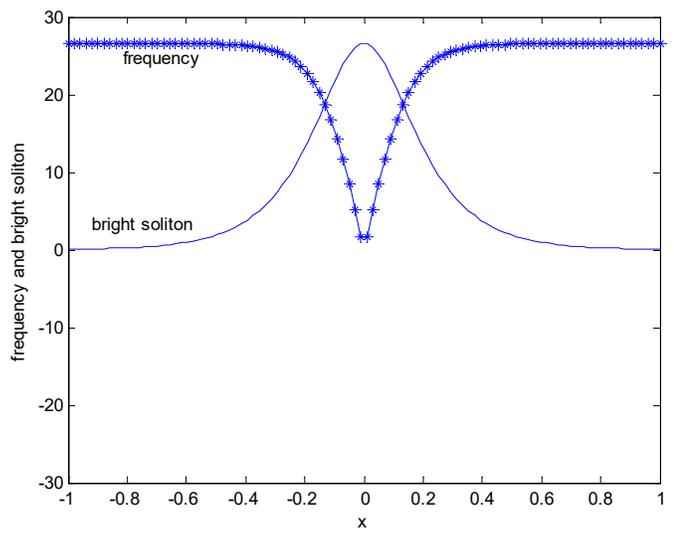

**Figure-3**

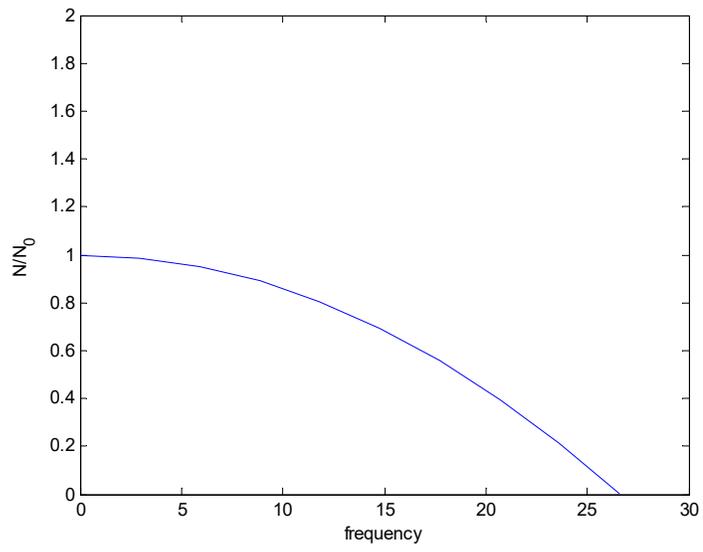

**Figure-4**